# Discovery of a cool planet of 5.5 Earth masses through gravitational microlensing*

J.-P. Beaulieu[1,4], D. P. Bennett[1,3,5], P. Fouqué[1,6], A. Williams[1,7], M. Dominik[1,8], U. G. Jørgensen[1,9], D. Kubas[1,10], A. Cassan[1,4], C. Coutures[1,11], J. Greenhill[1,12], K. Hill[1,12], J. Menzies[1,13], P.D. Sackett[1,14], M. Albrow[1,15], S. Brillant[1,10], J.A.R. Caldwell[1,16], J. J. Calitz[1,17], K. H. Cook[1,18], E. Corrales[1,4], M. Desort[1,4], S. Dieters[1,12], D. Dominis[1,19], J. Donatowicz[1,20], M. Hoffman[1,19], S. Kane[1,21], J.-B. Marquette[1,4], R. Martin[1,7], P. Meintjes[1,17], K. Pollard[1,15], K. Sahu[1,22], C. Vinter[1,9], J. Wambsganss[1,23], K. Woller[1,9], K. Horne[1,8], I. Steele[1,24], D. M. Bramich[1,8,24], M. Burgdorf[1,24], C. Snodgrass[1,25], M. Bode[1,24], A. Udalski[2,26], M.K. Szymański[2,26], M. Kubiak[2,26], T. Więckowski[2,26], G. Pietrzyński[2,26,27], I. Soszyński[2,26,27], O. Szewczyk[2,26], Ł. Wyrzykowski[2,26,28], B. Paczyński[2,29], F. Abe[3,30], I. A. Bond[3,31], T. R. Britton[3,15,32], A. C. Gilmore[3,15], J. B. Hearnshaw[3,15], Y. Itow[3,30], K. Kamiya[3,30], P. M. Kilmartin[3,15], A. V. Korpela[3,33], K. Masuda[3,30], Y. Matsubara[3,30], M. Motomura[3,30], Y. Muraki[3,30], S. Nakamura[3,30], C. Okada[3,30], K. Ohnishi[3,34], N. J. Rattenbury[3,28], T. Sako[3,30], S. Sato[3,35], M. Sasaki[3,30], T. Sekiguchi[3,30], D. J. Sullivan[3,33], P. J. Tristram[3,32], P. C. M. Yock[3,32], T. Yoshioka[3,30]

1. *PLANET/RoboNet Collaboration (http://planet.iap.fr, http://www.astro.livjm.ac.uk/RoboNet/),*
2. *OGLE Collaboration (http://ogle.astrouw.edu.pl),*
3. *MOA Collaboration (http://www.physics.auckland.ac.nz/moa),*
4. *Institut d'Astrophysique de Paris, CNRS, Université Pierre et Marie Curie UMR7095, 98bis Boulevard Arago, 75014 Paris, France,*
5. *University of Notre Dame, Department of Physics, Notre Dame, IN 46556-5670 United States of America*
6. *Observatoire Midi-Pyrénées, Laboratoire d'Astrophysique, UMR 5572, Université Paul Sabatier - Toulouse 3, 14, avenue Edouard Belin, 31400 Toulouse, France*
7. *Perth Observatory, Walnut Road, Bickley, Perth 6076, WA, Australia*
8. *Scottish Universities Physics Alliance, University of St Andrews, School of Physics & Astronomy, North Haugh, St Andrews KY16 9SS, United Kingdom*
9. *Niels Bohr Institutet, Astronomisk Observatorium, Juliane Maries Vej 30, 2100 København Ø, Denmark*
10. *European Southern Observatory, Casilla 19001, Santiago 19, Chile*
11. *CEA DAPNIA/SPP Saclay, 91191 Gif-sur-Yvette cedex, France*
12. *University of Tasmania, School of Mathematics and Physics, Private Bag 37, Hobart, TAS 7001, Australia*
13. *South African Astronomical Observatory, P.O. Box 9, Observatory 7935, South Africa*
14. *Research School of Astronomy and Astrophysics, Australian National University, Mt Stromlo Observatory, Weston Creek, ACT 2611, Australia*
15. *University of Canterbury, Department of Physics and Astronomy, Private Bag 4800 Christchurch 8020, New Zealand*
16. *McDonald Observatory, 16120 St Hwy Spur 78 #2, Fort Davis, TX 79734, United States of America*

17. *Boyden Observatory, University of the Free State, Department of Physics, PO Box 339 Bloemfontein 9300, South Africa*

18. *Lawrence Livermore National Laboratory, IGPP, P.O. Box 808 Livermore, CA 94551, United States of America*

19. *Universität Potsdam, Institut für Physik, Am Neuen Palais 10, 14469 Potsdam, Astrophysikalisches Institut Potsdam, An der Sternwarte 16, D-14482, Potsdam, Germany*

20. *Technische Universität Wien, Wiedner Hauptstr. 8 / 020 B. A.1040 Wien, Austria*

21. *Department of Astronomy, University of Florida, 211 Bryant Space Science Center, Gainesville, FL 32611-2055, United States of America*

22. *Space Telescope Science Institute, 3700 San Martin Drive, Baltimore, MD 21218, United States of America*

23. *Astronomisches Rechen-Institut (ARI), Zentrum für Astronomie, Universität Heidelberg, Mönchhofstr. 12–14, 69120 Heidelberg, Germany*

24. *Astrophysics Research Institute, Liverpool John Moores University, Twelve Quays House, Egerton Wharf, Birkenhead CH41 1LD, United Kingdom*

25. *Astronomy and Planetary Science Division, Department of Physics, Queen's University Belfast, Belfast, United Kingdom*

26. *Obserwatorium Astronomiczne Uniwersytetu Warszawskiego, Aleje Ujazdowskie 4, 00-478 Warszawa, Poland*

27. *Universidad de Concepcion, Departamento de Fisica, Casilla 160--C, Concepcion, Chile*

28. *Jodrell Bank Observatory, The University of Manchester, Macclesfield, Cheshire SK11 9DL, United Kingdom*

29. *Princeton University Observatory, Peyton Hall, Princeton, NJ 08544, United States of America*

30. *Solar-Terrestrial Environment Laboratory, Nagoya University, Nagoya, Japan*

31. *Institute for Information and Mathematical Sciences, Massey University, Auckland, New Zealand*

32. *Department of Physics, University of Auckland, Auckland, New Zealand*

33. *School of Chemical and Physical Sciences, Victoria University, Wellington, New Zealand*

34. *Nagano National College of Technology, Nagano, Japan*

35. *Department of Astrophysics, Faculty of Science, Nagoya University, Nagoya, Japan*

**In the favoured core-accretion model of formation of planetary systems, solid planetesimals accumulate to build up planetary cores, which then accrete nebular gas if they are sufficiently massive. Around M-dwarf stars (the most common stars in our Galaxy), this model favours the formation of Earth-mass ($M_{\oplus}$) to Neptune-mass planets with orbital radii of 1 to 10 astronomical units (AU), which is consistent with the small number of gas giant planets known to orbit M-dwarf host stars[1-4]. More than 170 extrasolar planets have been discovered with a wide range of masses and orbital periods, but planets of Neptune's mass or less have not hitherto been detected at separations of more than 0.15 AU from normal stars. Here we report the discovery of a $5.5^{+5.5}_{-2.7} M_{\oplus}$ planetary companion at a separation of $2.6^{+1.5}_{-0.6}$ AU from a $0.22^{+0.21}_{-0.11}$ solar mass ($M_{\odot}$) M-dwarf star. star, where $M_{\odot}$ refers to a solar mass. (We propose to name it OGLE-2005-BLG-390Lb, indicating a planetary mass companion to the lens star of the microlensing event.) The mass is lower than that of GJ876d (ref. 5), although the error bars overlap. Our detection suggests that such cool, sub-Neptune-mass planets may be more common than gas giant planets, as predicted by the core accretion theory.**

Gravitational microlensing events can reveal extrasolar planets orbiting the foreground lens stars if the light curves are measured frequently enough to characterize planetary light curve deviations with features lasting a few hours[6-9]. Microlensing is most sensitive to planets in Earth-to-Jupiter-like orbits with semi-major axes in the range 1-5 AU. The sensitivity of the microlensing method to low-mass planets is restricted by the finite angular size of the source stars[10,11], limiting detections to planets of a few Earth masses ($M_\oplus$) for giant source stars, but allowing the detection of planets as small as $0.1M_\oplus$ for main sequence source stars in the Galactic Bulge. The PLANET collaboration[12] maintains the high sampling rate required to detect low mass planets while monitoring the most promising of the > 500 microlensing events discovered annually by the OGLE collaboration, as well as events discovered by MOA. A decade of pioneering microlensing searches has resulted in the recent detections of two Jupiter-mass extrasolar planets[13,14] with orbital separations of a few AU by the combined observations of the OGLE, MOA, MicroFUN and PLANET collaborations. The absence of perturbations to stellar microlensing events can be used to constrain the presence of planetary lens companions. With large samples of events, upper limits on the frequency of Jupiter-mass planets have been placed over an **o**rbital range of 1-10 AU, down to Earth mass planets[15-17] for the most common stars of our galaxy.

On 11 July 2005, the OGLE Early Warning System[18] announced the microlensing event OGLE-2005-BLG-390 ($\alpha$ = 17:54:19.2, $\delta$ = -30:22:38, J2000) with a relatively bright clump giant as a source star. Subsequently, PLANET, OGLE and MOA monitored it with their different telescopes. After peaking at a maximum magnification of $A_{\rm max} = 3.0$ on July 31, 2005, a short duration deviation from a single lens light curve was detected on 9 August 2005 by PLANET. As described below, this deviation was due to a low-mass planet orbiting the lens star.

From analysis of colour-magnitude diagrams, we derive the following reddening-corrected colours and magnitudes for the source star: $(V-I)_0 = 0.85$, $I_0 = 14.25$ and $(V-K)_0 = 1.9$. We used the surface brightness relation[20] linking the emerging flux per solid angle of a light-emitting body to its colour, calibrated by interferometric observations, to derive an angular radius of 5.25 ± 0.73 μas, which corresponds to a source radius of $9.6 \pm 1.3 R_\odot$ if the source star is at a distance of 8.5 kpc. From its colours, the star is a 5200 K giant, which corresponds to G4 III spectral type.

Figure 1 shows our photometric data for microlensing event OGLE 2005-BLG-390 and the best planetary binary lens model. The best-fit model has $\chi^2 = 562.26$ for 650 data points, 7 lens parameters, and 12 flux normalization parameters, for a total of 631 degrees or freedom. Model length parameters in Table 1 are expressed in units of the Einstein ring radius $R_E$ (typically ~2 AU for a Galactic Bulge system), the size of the ring image that would be seen in case of perfect lens-source alignment. In modelling the light curve, we adopted linear limb darkening laws[21] with $\Gamma_I = 0.538$ and $\Gamma_R = 0.626$, appropriate for this G4III giant source star, to describe the centre-to-limb variation of the intensity profile in the I and R bands. Four different binary lens modelling codes were used to confirm that the model we present is the only acceptable model for the observed light curve. The best alternative model is one with a large flux-ratio binary source with a single lens, which has gross features that are similar to a planetary microlensing event[22]. However, as shown in Figure 1, this model fails to account for the PLANET-Perth, PLANET-Danish and OGLE measurements near the end of the planetary deviation, and it is formally excluded by $\Delta\chi^2 = 46.25$ with 1 fewer model parameters.

The planet is designated OGLE-2005-BLG-390Lb, where the "Lb" suffix indicates the secondary component of the lens system with a planetary mass ratio. The microlensing fit determines directly only the planet-to-star mass ratio, $q = 7.6 \pm 0.7 \times 10^{-5}$ and the projected separation in units of $R_E$, $d = 1.610 \pm 0.008$. Although the planet and star masses are not directly determined for planetary microlensing events, we can derive their probabilities densities. We have performed a Bayesian analysis[23] employing the Galactic models and mass functions described in refs. [11,23]. We averaged over the distances and velocities of the lens and source stars, subject to the constraints due to the angular diameter of the source and the measured parameters given in Table 1. This analysis gives a 95% probability that the planetary host star is a main sequence star, a 4% probability that it is a white dwarf, and a probability of < 1% that it is a neutron star or black hole. The host star and planet parameter probability densities for a main sequence lens star are shown in Figure 2 for the Galactic model used in ref. [23]. The medians of the lens parameter probability distributions yield a companion mass of $5.5^{+5.5}_{-2.7} M_\oplus$ and an orbital separation of $2.6^{+1.5}_{-0.6}$ AU from the $0.22^{+0.21}_{-0.11} M_\odot$ lens star, which is located at a distance of $D_L = 6.6 \pm 1.0$ kpc, where the error bars indicate the central 68% confidence interval. These median parameters imply that the planet receives only 0.1% of the radiation from its host star that the Earth receives from the Sun, so the likely surface temperature of the planet is ~50 K, similar to the temperatures of Neptune and Pluto.

The parameters of this event are near the limits of microlensing planet detectability for a giant source star. The separation of $d = 1.61$ is near the outer edge of the so-called lensing zone[7], which has the highest planet detection probability, and the planet's mass is about a factor 2 above the detection limit set by the finite size of the source star. Planets with $q > 10^{-3}$ and $d \approx 1$ are much easier to detect, and so it may be that the parameters of OGLE 2005-BLG-390Lb represent a more common type of planet. This can be quantified by simulating planetary light curves with different values of $q$ and $\theta$, but the remaining parameters fixed to the values for the 3 known microlensing planets. We find that the probability of detecting a $q \approx 4\text{-}7 \times 10^{-3}$ planet, like the first two microlens planets[13,14], is ~50 times larger than the probability of detecting a $q = 7.6 \times 10^{-5}$ planet like OGLE 2005-BLG-390Lb. This suggests that, at the orbital separations probed by microlensing, sub-Neptune mass planets are significantly more common than large gas giants around the most common stars in our Galaxy. Similarly, the first detection of a sub-Neptune mass planet at the outer edge of the "lensing zone" provides a hint that these sub-Neptune mass planets may tend to reside in orbits with semi-major axes $a > 2$ AU.

The core-accretion model of planet formation predicts that rocky/icy 5-15 Earth-mass planets orbiting their host stars at 1-10 AU are much more common than Jupiter-mass planets, and this prediction is consistent with the small fraction of M-dwarfs with planets detected by radial velocities[3,5] and with previous limits from microlensing[15]. Our discovery of such a low-mass planet by gravitational microlensing lends further support to this model, but more detections of similar and lower mass planets over a wide range of orbits are clearly needed. Planets with separations of ~0.1 AU will be detected routinely by the radial velocity method or space observations of planetary transits in the coming years[24,25,26,27], but the best chance to increase our understanding of such planets over orbits of 1-10 AU in the next 5-10 years is by future interferometer programs[28] and more advanced microlensing surveys[11,29,30].

## REFERENCES

1. Safronov, V. 1969, Evolution of the Protoplanetary Cloud and Formation of the Earth and Planets (Moscow: Nauka)
2. Wetherill, G. W. Formation of the terrestrial planets *Ann. Rev. Astron. Astrophys.* **18**, 77-113 (1980).
3. Laughlin, G., Bodenheimer, P., & Adams, F. C. The Core Accretion Model Predicts Few Jovian-Mass Planets Orbiting Red Dwarfs. *Astrophys. J.* **612**, L73-L76 (2004).
4. Ida, S., & Lin, D. N. C. Toward a Deterministic Model of Planetary Formation. II. The Formation and Retention of Gas Giant Planets around Stars with a Range of Metallicities. *Astrophys. J.* **616**, 567-572 (2004).
5. Rivera E., et al. A ~ 7.5 Earth-Mass Planet Orbiting the Nearby Star, GJ 876 *Astrophys. J.* in press (2005).
6. Mao, S., & Paczynski, B. Gravitational microlensing by double stars and planetary systems *Astrophys. J.* **374**, L37-L40 (1991).
7. Gould, A., & Loeb, A. Discovering planetary systems through gravitational Microlenses. *Astrophys. J.* **396**, 104-114 (1992).
8. Wambsganss, J. Discovering Galactic planets by gravitational microlensing: magnification patterns and light curves *Mon. Not. R. Soc.* **284**, 172-188 (1997).
9. Griest, K., & Safizadeh, N. The Use of High-Magnification Microlensing Events in Discovering Extrasolar Planets *Astrophys. J.* **500**, 37 (1998).
10. Bennett, D. P., & Rhie, S. H. Detecting Earth-Mass Planets with Gravitational Microlensing *Astrophys. J.* **472**, 660 (1996).
11. Bennett, D. P., & Rhie, S. H. Simulation of a Space-based Microlensing Survey for Terrestrial Extrasolar Planets *Astrophys. J.* **574**, 985-1003 (2002).
12. Albrow, M., et al. The 1995 Pilot Campaign of PLANET: Searching for Microlensing Anomalies through Precise, Rapid, Round-the-Clock Monitoring *Astrophys. J.* **509**, 687-702 (1998).
13. Bond, I. A., et al. OGLE 2003-BLG-235/MOA 2003-BLG-53: A Planetary Microlensing Event *Astrophys. J.* **606**, L155-L158 (2004).
14. Udalski, A., et al. A Jovian-Mass Planet in Microlensing Event OGLE-2005-BLG-071 *Astrophys. J.* **628**, L109-L112 (2005).
15. Gaudi, B. S., et al. Microlensing Constraints on the Frequency of Jupiter-Mass Companions: Analysis of 5 Years of PLANET Photometry *Astrophys. J.* **566**, 463-499 (2002).
16. Abe, F., et al. Search for Low-Mass Exoplanets by Gravitational Microlensing at High Magnification, *Science*, **305**, 1264-1267 (2004).

17. Dong, S., et al. Planetary Detection Efficiency of the Magnification 3000 Microlensing Event OGLE-2004-BLG-343 *Astrophys. J.* submitted (2005), astro-ph/0507079.

18. Udalski, A. The Optical Gravitational Lensing Experiment. Real Time Data Analysis Systems in the OGLE-III Survey. *Acta Astronomica* **53**, 291-305 (2003).

19. Alard, C. Image subtraction using a space-varying kernel. *Astron. Astrophys. Suppl.* **144**, 363-370 (2000).

20. Kervella, P., et al. Cepheid distances from infrared long-baseline interferometry. III. Calibration of the surface brightness-color relations. *Astron. Astrophys.* **428**, 587-593 (2004).

21. Claret, A., Diaz-Cordoves, J., & Gimenez, A. Linear and non-linear limb-darkening coefficients for the photometric bands R I J H K. *Astron. Astrophys. Suppl.* **114** , 247 (1995).

22. Gaudi, B. S., A. Distinguishing Between Binary-Source and Planetary Microlensing Perturbations. *Astrophys. J.* **506**, 533 (1998).

23. Dominik, M., Stochastical distributions of lens and source properties for observed galactic microlensing events. Submitted to MNRAS (2005), astro-ph/ 0507540.

24. Vogt, S. S., et al. Five New Multicomponent Planetary Systems *Astrophys. J.* **632**, 638-658 (2005).

25. Mayor, M., et al. The CORALIE survey for southern extrasolar planets. XII. Orbital solutions for 16 extrasolar planets discovered with CORALIE. *Astron. Astrophys.* **415**, 391-402 (2004).

26. Borucki, W. et al. in Second Eddington Workshop: Stellar structure and habitable planet finding, ed. F. Favata, S. Aigrain, & A. Wilson, *ESA SP*-**538**, 177 (2004).

27. Moutou, C., et al. Comparative blind test of five planetary transit detection algorithms on realistic synthetic light curves *Astron. Astrophys.* **437**, 355-368 (2005).

28. Sozzeti, A., et al., Narrow-Angle Astrometry with the Space Interferometry Mission: The Search for Extrasolar Planets. I. Detection and Characterization of Single Planets, *Pub. Astr. Soc. Pac.* **114**, 1173-1196 (2002).

29. Bennett, D. P. The Detection of Terrestrial Planets via Gravitational Microlensing: Space vs. Ground-based Surveys *ASP Conf. Ser. : Extrasolar Planets: Today and Tomorrow* **321**, 59 (2004).

30. Beaulieu, J. P., et al. PLANET III: searching for Earth-mass planets via microlensing from Dome C? *EAS Publications Series* **14**, 297-302 (2005).

**Acknowledgements** PLANET is grateful to the observatories that support our science (European Southern Observatory, Canopus, Perth, South African Astronomical Observatory, Boyden, Faulkes North) and to the ESO people in La Silla for their help to maintain and operate the Danish telescope. Support for the PLANET project was provided by CNRS, NASA, NSF, LLNL/NNSA/DOE, PNP, David Warren, DFG, IDA and SNF. RoboNet is funded by the UK PPARC and the FTN was supported by the Dill Faulkes Educational Trust. Support for the OGLE project, conducted at Las Campanas Observatory operated by the Carnegie Institution of Washington, was provided by the Polish Ministry of Science, the Foundation for Polish Science, NSF and NASA. The MOA collaboration is supported by MEXT and JSPS of Japan, and the Marsden Fund of New Zealand.

**Author Information** The photometric data set is available at planet.iap.fr and ogle.astrouw.edu.pl Reprint and permissions information is available at npg.nature.com/reprintsandpermissions. . The authors declare no competing financial interests. Correspondence and requests for materials should be addressed to JP B (beaulieu@iap.fr) or D.P.B. (bennett@nd.edu).

| Microlensing Fit Parameters | |
|---|---|
| $d$ = (star-planet separation) $/R_E$ | 1.610 ± 0.008 |
| $q$ = planet:star mass ratio | $(7.6 \pm 0.7) \times 10^{-5}$ |
| $u_0$ = (closest approach)$/R_E$ | 0.359 ± 0.005 |
| $t_E$ = Einstein ring radius crossing time | 11.03±0.11 days |
| $t_0$ = time of closest approach | 31.231 ± 0.005 July 2005 UT |
| $t_*$ = source star radius crossing time | 0.282 ± 0.010 days |
| $\theta$ = angle of source motion | 2.756 ± 0.003 rad |

Table 1: The parameters for the best binary lens model for the OGLE 2005-BLG-390 microlensing event light curve are shown with their 1-σ uncertainties. A number of these parameters are scaled to the Einstein ring radius, which is given by $R_E = 2\sqrt{GMD_L(D_S - D_L)/(c^2 D_S)}$, where $D_L$ and $D_S$ are the lens and source distances, respectively. $\theta$ is the angle of source motion with respect to the lens axis.

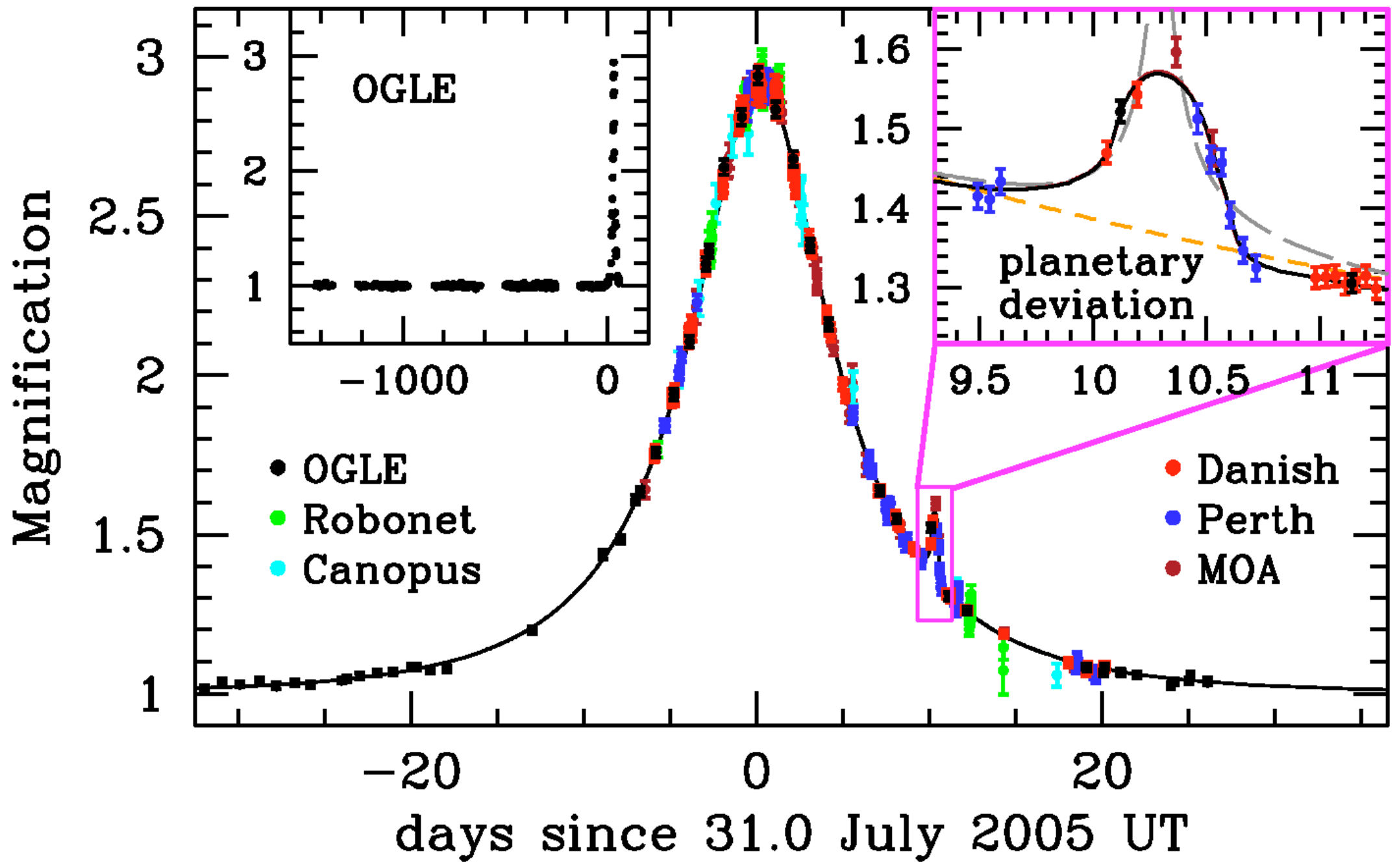

Figure 1 : The observed light curve of the OGLE-2005-BLG-390 microlensing event and best fit model plotted as a function of time. The data set consists of 650 data points from PLANET Danish (ESO La Silla, red points), PLANET Perth (blue), PLANET Canopus (Hobart, cyan), RoboNet Faulkes North (Hawaii, green), OGLE (Las Campanas, black), MOA (Mt John Observatory, brown). This photometric monitoring was done in the I band (with the exception of Faulkes R band data and MOA custom red passband) and real-time data reduction was performed with the different OGLE, PLANET and MOA data reduction pipelines. Danish and Perth data were finally reduced by the image subtraction technique[19] with the OGLE pipeline. The top left inset shows the OGLE light curve extending over the previous 4 years, whereas the top right one shows a zoom of the planetary deviation, covering a time interval of 1.5 days. The solid curve is the best binary lens model described in the text with a planet-to-star mass ratio of $q = 7.6 \pm 0.7 \times 10^{-5}$, and a projected separation $d = 1.610 \pm 0.008\ R_E$ (where $R_E$ is the Einstein ring radius). The dashed grey curve is the best binary source model that is rejected by the data, while the dashed orange line is the best single lens model.

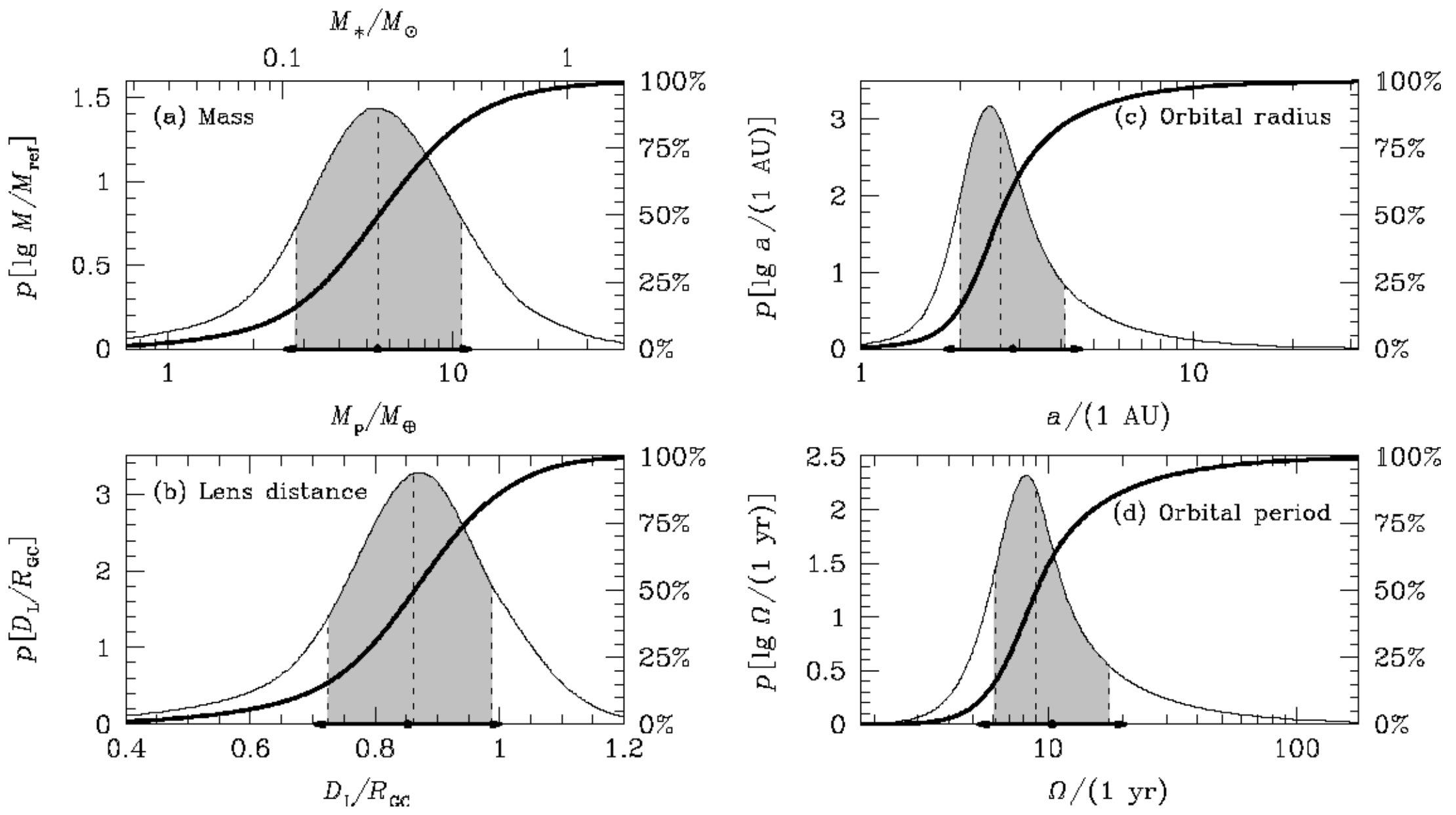

Figure 2 : Bayesian probability densities for the properties of the planet and its host star. The individual panels show the masses of the lens star and its planet (a), their distance from the observer (b), the three-dimensional separation or semi-major axis of an assumed circular planetary orbit (c) and the orbital period of the planet (d). The bold, curved line in each panel is the cumulative distribution, with the percentiles listed on the right. The dashed vertical lines indicate the medians, and the shading indicates the central 68.3% confidence intervals, while dots and arrows on the abscissa mark the expectation value and standard deviation. All estimates follow from a Bayesian analysis assuming a standard model for the disk and bulge population of the Milky Way and the stellar mass function of ref. [23], and a prior for the source distance $D_S$=1.05 ±0.25 $R_{GC}$ (where $R_{GC}$= 7.62 ± 0.32 kpc for the Galactic Centre distance). The medians of these distributions yield a $5.5^{+5.5}_{-2.7}$ Earth mass planetary companion at a separation of $2.6^{+1.5}_{-0.6}$ AU from a $0.22^{+0.21}_{-0.11} M_{\odot}$ Galactic Bulge M-dwarf at a distance of 6.6 ± 1.0 kpc from the Sun. The median planetary period is $9^{+9}_{-3}$ years. The logarithmic means of these probability distributions (which obey Kepler's third law) are a separation of 2.9 AU, a period of 10.4 years, and masses of $0.22M_{\odot}$ and $5.5M_{\oplus}$ for the star and planet, respectively. In each plot, the independent variable for the probability density is listed within square brackets. The distribution of planet-star mass ratio was taken to be independent of the stellar mass, and a uniform prior was assumed for the planet-star separation distribution.